\documentclass[conference]{IEEEtran}
\IEEEoverridecommandlockouts
\usepackage{cite}
\usepackage{amsmath,amssymb,amsfonts}
\usepackage{algorithmic}
\usepackage{graphicx}
\usepackage{textcomp}
\usepackage{xcolor}
\usepackage{url}

\def\BibTeX{{\rm B\kern-.05em{\sc i\kern-.025em b}\kern-.08em
    T\kern-.1667em\lower.7ex\hbox{E}\kern-.125emX}}
\begin{document}

\title{Using Ensemble Inference to Improve Recall of Clone Detection}

\author{
\IEEEauthorblockN{Gul Aftab Ahmed}
\IEEEauthorblockA{\textit{Lero Research Centre}\\
\textit{Trinity College Dublin} \\
Dublin, Ireland \\
ahmedga@tcd.ie}
\and
\IEEEauthorblockN{James Vincent Patten}
\IEEEauthorblockA{\textit{Dept. of Computer Science and} \\
\textit{Information Systems} \\
\textit{University of Limerick}\\
Limerick, Ireland \\
james.patten@lero.ie}
\and
\IEEEauthorblockN{Yuanhua Han}
\IEEEauthorblockA{\textit{WN Digital IPD and} \\
\textit{Trustworthiness Enabling} \\
\textit{Huawei Technologies Co., Ltd.}\\
Xi'an, Shaanxi, China \\
hanyuanhua2@huawei.com}
\and
\IEEEauthorblockN{Guoxian Lu}
\IEEEauthorblockA{\textit{WN Digital IPD and} \\
\textit{Trustworthiness Enabling} \\
\textit{Huawei Technologies Co., Ltd.}\\
Shanghai, China \\
luguoxian@huawei.com}
\and
\IEEEauthorblockN{David Gregg}
\IEEEauthorblockA{\textit{Lero Research Centre}\\
\textit{Trinity College Dublin} \\
Dublin, Ireland \\
david.gregg@tcd.ie}
\and
\IEEEauthorblockN{Jim Buckley}
\IEEEauthorblockA{\textit{Dept. of Computer Science and} \\
\textit{Information Systems} \\
\textit{University of Limerick}\\
Limerick, Ireland \\
jim.buckley@ul.ie}
\and
\IEEEauthorblockN{Muslim Chochlov}
\IEEEauthorblockA{\textit{Dept. of Computer Science and} \\
\textit{Information Systems} \\
\textit{University of Limerick}\\
Limerick, Ireland \\
muslim.chochlov@ul.ie}

}

\hyphenation{SourcererCC}

\maketitle

\begin{abstract}
Large-scale source-code clone detection is a challenging task. In our previous work, we proposed an approach (SSCD) that leverages artificial neural networks and approximates nearest neighbour search to effectively and efficiently locate clones in large-scale bodies of code, in a time-efficient manner. However, our literature review suggests that the relative efficacy of differing neural network models has not been assessed in the context of large-scale clone detection approaches. In this work, we aim to assess several such models individually, in terms of their potential to maximize recall, while preserving a high level of precision during clone detection. We investigate if ensemble inference (in this case, using the results of more than one of these neural network models in combination) can further assist in this task.

To assess this, we employed four state-of-the-art neural network models and evaluated them individually/in combination. The results, on an illustrative dataset of approximately 500K lines of C/C++ code, suggest that  ensemble inference outperforms individual models in all trialled cases, when recall is concerned. Of individual models, the ADA model (belonging to the ChatGPT family of models) has the best performance. However commercial companies may not be prepared to hand their proprietary source code over to the cloud, as required by that approach. Consequently, they may be more interested in an ensemble-combination of CodeBERT-based and CodeT5 models, resulting in similar (if slightly lesser) recall and precision results.

\end{abstract}

\section*{Note to Practitioners}
This is the accepted version of the paper submitted to the International Workshop on Software Clones (IWSC 2023). The final version should be accessible at \url{https://doi.org/10.1109/IWSC60764.2023.00010}.
\section{Introduction}

Clone detection is the process of locating textually or functionally exact or similar pieces of code\cite{Rattan2013}. The reasons for performing clone detection vary, but can involve the need to address software maintenance issues (where, for example, a vulnerability found in the clone-source should be alerted to the clone-destination) and licensing issues (where code from one licensing regime has historically, and inappropriately, been cloned into a system with another licensing regime) among other tasks. 

Locating clones manually in large pieces of code is effort-intensive and so, over the past few decades, many clone detection approaches and their implementing tools were proposed, showing mixed results \cite{Rattan2013, Ain2019}, particularly with respect to type three and type four clones where the textual-code match is less exact and so techniques need to be more sophisticated \cite{Rattan2013}. 

More recently, artificial intelligence (AI) models were utilized and reported state-of-the-art results in pairwise clone detection tasks \cite{Feng2020, Guo2020}. However, these AI-based approaches do not address clone detection at scale, which is often needed in real-world settings.  This is because the pair-wise comparison of code-segments results in combinatorial effort, as the scale of the systems involved grows.

In our previous work, a novel approach (SSCD) was proposed that addresses this issue: it allows for AI-based, clone detection at scale, based on a global, nearest-neighbour comparison of  code-segment vectors encoded by a fine-tuned, CodeBERT-based ANN \cite{chochlov2022using}. When applied to our partner company's C and C++ datasets\footnote{\url{https://github.com/SFI-Lero/SSCD}}, and to an open-source 320 million LOC BigCloneBench (BCB) dataset\cite{Svajlenko2017a}, SSCD was both effective and efficient, outperforming baseline approaches such as SAGA and SourcererCC \cite{chochlov2022using}.

Extending this work, research has focused on approaches to improve the clone-detection recall (focusing on the underlying AI-based inference component), while also maintaining precision. The rationale for higher recall is to find as many clones as reasonably possible, for example, to assist with vulnerability detection. Improving accuracy of AI-based artificial neural network (ANN) models is a non-trivial task and several strategies can be adopted. For example:

\begin{itemize}
    \item Designing new ANNs \cite{vaswani2017attention}, or re-designing and changing their architecture, parameters, or inputs \cite{Guo2020}
    \item Pre-training and fine-tuning ANNs \cite{Devlin2018}
    \item      Incorporating alternative ANNs
    \item Ensemble learning or inference: using multiple ANNs, designed and trained towards a similar task, working together \cite{ghosh2021using}
\end{itemize}

Our literature review suggests that differing ANNs have not yet been assessed for clone detection at scale. Hence, in the work reported here, we focus on the two latter-most strategies: assessing alternative, state-of-the-art ANNs for clone detection and ensemble inference. These strategies were selected because intuitively state-of-the-art models would seem to promise improved accuracy of downstream tasks \cite{ghosh2021using} with a low practical-application barrier, which can be valuable in industrial research. Likewise, ensembles of state-of-the-art ANNs would seem likely to increase efficacy further, at least in terms of recall. Here, four existing transformer-architecture based \cite{vaswani2017attention} ANNs were used: 

\begin{itemize}
    \item CodeBERT (fine-tuned for clone detection)\footnote{\url{https://huggingface.co/mchochlov/codebert-base-cd-ft}} (CBF)\cite{Feng2020, chochlov2022using};
    \item GraphCodeBERT (GCB) \cite{Guo2020}; 
    \item CodeT5 (CT5) \cite{Wang2021}, and 
    \item OpenAI ADA version 2 (ADA) GPT-3 model \cite{openai}.
\end{itemize} 

These models were selected because of reported state-of-the-art results in source code tasks (including clone detection). Yet these models are distinct enough in their characteristics to suggest possible recall improvement when used in combination, leveraging their individual strengths and capabilities. 

To enhance performance, first, averaging, pipelining, and stacking were considered as ensembling strategies. For example, using the stacking, the embedding outputs of individual models were combined to a higher-level ``meta-model'' embedding. However, the strategy did not yield effective. 

Here in our adopted approach, we determine the results for each ANN individually and then combine the results of these models in all possible combinations (ensemble inferencing) towards improving the accuracy of the downstream task: clone detection. This was done to address the following research question (RQ): 
\begin{enumerate}
    \item \textbf{How do differing ANNs affect the recall and overall efficacy of a large-scale, AI-based, clone-detection approach like SSCD?}
    \item \textbf{How does ensemble inference, based on combining these differing ANNs,  affect the recall and overall efficacy of a large-scale, AI-based, clone detection approach like SSCD?}
\end{enumerate}

The contributions of this work are following:

\begin{itemize}
    \item It demonstrates that, by using ensemble inference (combining the outputs of the four selected ANNs), recall with respect to clone detection is improved, as compared to if these models were used individually. Additionally, and somewhat surprisingly, the decrease in precision when using these ensembles is not as drastic as one might expect.
    \item It demonstrates significant difference in recall and precision performance between four different, state-of-the-art source-code-targeted ANNs, when applied individually;
    \item In doing so, it clearly identifies the ChatGPT ADA model as most appropriate for high-accuracy clone detection;
    
\end{itemize}

\begin{table*}[htbp]
\renewcommand{\arraystretch}{1.5}
\caption{Characteristics of Transformer-based ANNs Used in This Work}
\label{tbl:ann_details}
\centering
\resizebox{\textwidth}{!}{\begin{tabular}{l|llrrlr}
\hline
\multicolumn{1}{c|}{Model} & \multicolumn{1}{c}{\begin{tabular}[c]{@{}c@{}}Training\\ Status\end{tabular}} & \multicolumn{1}{c}{Input}                                               & \multicolumn{1}{c}{Year} & \multicolumn{1}{c}{\# Parameters} & \multicolumn{1}{c}{PL used for training}                                                             & \multicolumn{1}{c}{\begin{tabular}[c]{@{}c@{}}Embedding \\ Size\end{tabular}} \\ \hline
ADA                        & N/A                                                                           & NL/PL                                                                   & 2022                     & N/A                               & N/A                                                                                                  & 1536                                                                          \\ \hline
CT5                        & PT                                                                            & \begin{tabular}[c]{@{}l@{}}NL/PL \\ (focus on identifiers)\end{tabular} & 2021                     & 223M                              & Ruby, JavaScript, Go, Python, Java, PHP, C, C\#                                                      & 768                                                                           \\ \hline
CBF                        & PT/FT                                                                         & NL/PL                                                                   & 2022                     & 125M                              & \begin{tabular}[c]{@{}l@{}}PT: Go, Java, JavaScript, PHP, Python, Ruby\\ FT: Java (BCB)\end{tabular} & 768                                                                           \\ \hline
GCB                        & PT                                                                            & NL/PL/ data flow graph                                                  & 2020                     & 125M                              & PT: Go, Java, JavaScript, PHP, Python, Ruby                                                          & 768                                                                           \\ \hline
\end{tabular}
}
\end{table*}

The rest of the paper is organized as follows: in Section~\ref{sec:background}, ANNs are introduced and the selected ANNs are discussed in detail, with their differences explained. Also the section, briefly discusses SSCD and changes to SSCD that were needed to run the "alternative ANNs" experiment that is the focus of this paper. Section~\ref{sec:methodology} presents the experimental methodology and Section~\ref{sec:results} talks to the results of that experiment. Section~\ref{sec:threats} touches on threats to the validity of this experiment and Section~\ref{sec:conclusions} summarizes conclusions, giving directions for future work.

\section{Background}
\label{sec:background}


In the last decade new, deep neural networks with many layers have become the most accurate computer-based solution to a growing number of problems. The first big successes were in the area of image classification and processing, where ANNs give more accurate results than classical computer vision. More recently, deep neural network models have also greatly improved the capacity and accuracy of text and language processing. Recurrent neural networks (RNN), where connections between the nodes in the neural network can form a cycle, and thus allow output from nodes to affect subsequent input to those nodes, have proven especially accurate for text processing, but they require a great deal of time to train. 

Transformer-based models (described below), such as Bidirectional Encoder Representations from Transformers (BERT), allow greater parallelism in the training compared to RNNs, meaning that much larger models can be trained. More recently, generative pre-trained transformer (GPT) models with billions of parameters have achieved even higher accuracy in language processing and generation. These models have been trained to operate both on natural language and on programming languages.

\subsection{Transformer based Models Used in This Work}
ANNs used in this work, and their characteristics, are shown in Table~\ref{tbl:ann_details}. Abbreviations used in the table (and further in the text) are as follows: PT stands for ``pre-trained''; where we use the existing model without modification. FT stands for ``fine-tuned'', meaning that we have taken the original pre-trained model and done additional fine-tune training using clone detection training data. The input to the model may be ``natural language'' (NL), or ``programming language'' (PL). In the ``\# Parameters'' column, M stands for ``millions''. OpenAI ADA version 2 (ADA) is a proprietary GPT-3 based model developed by OpenAI \cite{openai}. Some of its parameters are unknown and therefore there is a N/A designation in that cell, as with the programming language(s) used for training.

\textbf{Common} to all ANNs used in this work is whats known as a transformer architecture \cite{vaswani2017attention}. Characteristic of this architecture is the usage of attention layer(s) that allow for learning of a word/token representation from its context. The encoder part of these models takes source code as an input and, using the tokens' contexts, returns an associated numeric representation (embedding). All the models were pre-trained: they learned a certain generic language model making them suitable for a broad spectrum of NL/PL tasks. 

Similarly, all of the models are fairly recent (having appeared in the past 3 years) and have showed state-of-the-art results in natural and programming language tasks particularly, suggesting their suitability for the clone detection task probed here \cite{chochlov2022using, openai, Guo2020, Wang2021}.

\textbf{The differences} in these models can be found in their degree of training, input details, their number of parameters, the types of programming languages used for training, and the size of generated embeddings (see Table~\ref{tbl:ann_details}). For example, in addition to pre-training, CBF was fine-tuned (further trained) towards clone detection specifically (see our previous work for more details \cite{chochlov2022using}). The input to these models is a mix of NL (e.g. comments) and PL, as found in source code. Unlike other models, CT5 focuses on identifiers in the source code and GCB adds structural information (data flow), derived from source code. CT5 has twice as many parameters as CBF/GCB. The set of programming languages used for training the ANNs includes Ruby, Javascript, Go, Python, Java, and PHP, but is individual to each model. In addition to this set, CT5 was also trained on C (one of the languages used in our industrial partner's codebase) and C\#. Finally, the embedding size in ADA is twice as large as the rest of the models.

The nature of these models (usage directed towards NL/PL tasks) suggests their suitability for clone detection in isolation. Yet their differences suggest they are heterogeneous enough to be also leveraged together to improve the recall of clone detection: in clone-instances where one model fails another can excel, thus making them more effective cumulatively. The reasoning for relying on the complementary strengths of the chosen ANNs specifically are: 

\begin{itemize}
    \item Fine-tuning with respect to clone detection should improve effectiveness of ANNs in downstream tasks \cite{Devlin2018};
    \item The inclusion of structural information could improve clone detection \cite{Jiang2007};
    \item  The number of parameters in ANNs could improve their effectiveness \cite{neelakantan2022text}. 
\end{itemize}

We also assume that using the same PL for training and inference (e.g. C language in CT5) can be beneficial. For example, in our previous work, we have fine-tuned CBF on Java-language code-clones, but used that model with C/C++ datasets \cite{chochlov2022using}. 

Finally, the ADA model belongs to either the GPT3 or GPT3.5 model family \cite{brown2020language, openai} that recently showed state-of-the-art results in NL/PL tasks and led to the success of ChatGPT (for example \cite{kasneci2023chatgpt}).  These factors, and specifically the individual characteristics that suggested individual ANN suitability for clone detection, prompted their selection for inclusion in this study.


\subsection{Scalable clone detection}
This section provides an overview of the architecture of  SSCD, an approach which is primarily motivated by the need for a more efficient and scalable approach to code clone detection, especially within large-scale code repositories. Several ANN based approaches  ASTNN \cite{Zhang2019}, CodeBERT \cite{Feng2020}, and GraphCodeBERT \cite{Guo2021}  to  clone detection  have improved the accuracy of detection of T3/T4 type clones but they rely on pairwise comparison of code fragments and thus can not scale to large code bases  due to the quadratic complexity of said pairwise comparisons.
In light of these challenges, SSCD was developed to generate numerical representations or 'embeddings' for each code fragment using artificial neural networks (transformer architecture). By transitioning to this embedding-based approach, SSCD can leverage efficient approximate k-nearest neighbour (k-NN) algorithms for clone detection, instead of performing exhaustive pairwise comparisons, finding the 'k' nearest neighbours for each embedding in logarithmic time complexity \cite{chochlov2022using}.

 The underlying models used in this process are fine-tuned versions of pretrained models—CodeBERT and GraphCodeBERT—which are adapted by adding a pooling layer to generate a 768-dimensional vector representation for each code fragment. For the CodeBERT-fine-tuned model, only textual information from the source code is used to generate vectors, whereas the GraphCodeBERT-fine-tuned model also includes structural information to enhance the richness of the representation.

The tool ranks the results based on their cosine similarity, producing a list of similar code fragments for each piece of source code. The generation of this ranked list is controlled by two parameters: a cosine similarity threshold and a 'topN' parameter that determines the number of results returned.

In summary, the SSCD integrates embedding generation, efficient search techniques, and result ranking, thus providing a scalable and effective solution for clone detection in large codebases.

\subsection{Changes to SSCD in This Work}

\begin{figure}[htbp]
\includegraphics[width=\columnwidth]{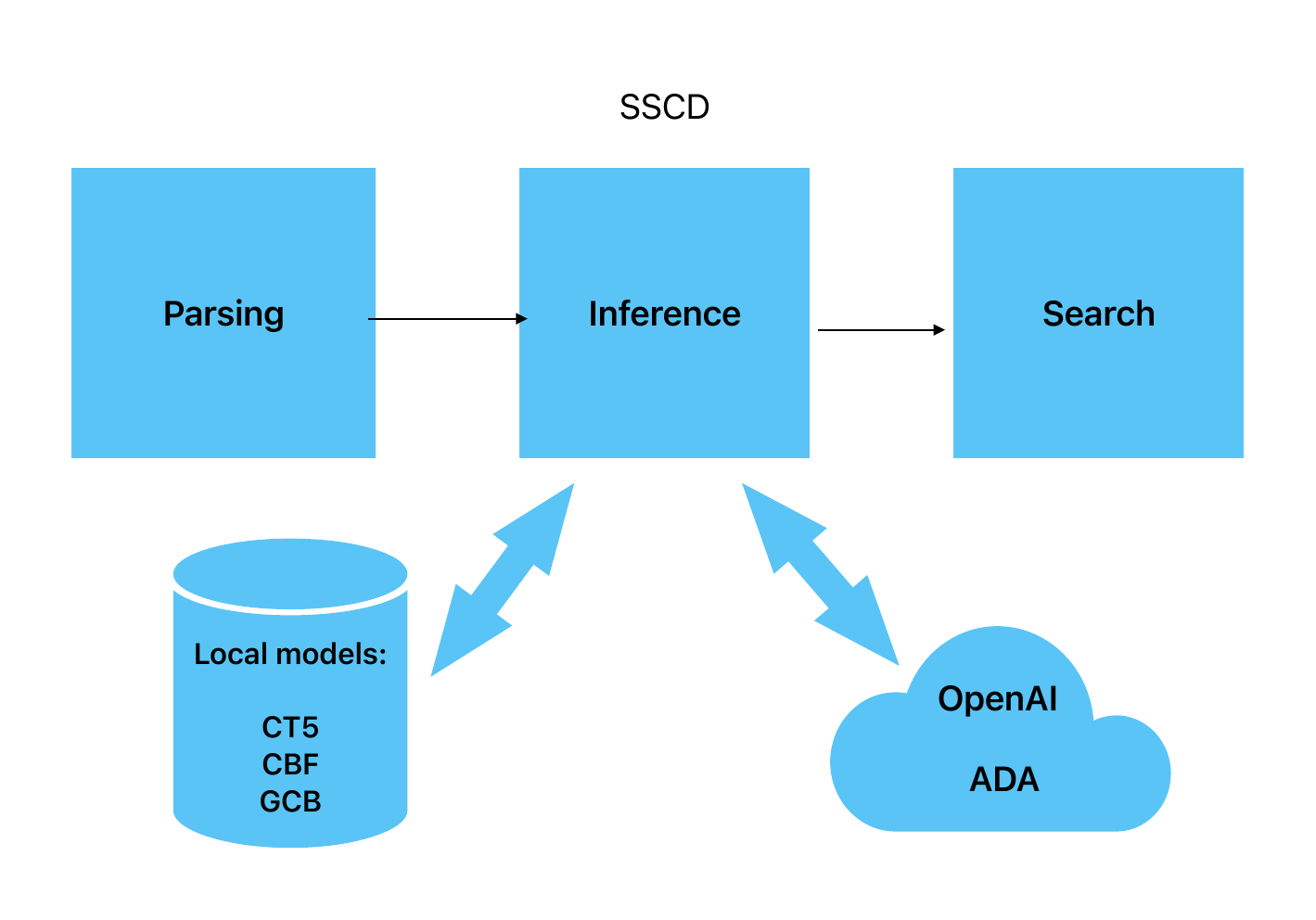}
\caption{The Updated Schema of SSCD}
\label{fig:sscd}
\end{figure}

As can be seen in Figure~\ref{fig:sscd}, SSCD has 3 major components: parsing, inference (employing ANNs), and search (relying on k approximate nearest neighbour (kANN) for scalability \cite{Malkov2018}). The input to the approach is source code and the output is a set of nearest neighbour clone candidates. The approach operates at function-level granularity and its implementing tool currently supports Java, C, and C++ languages.

In our previous work, we relied on locally available models such as CBF and GCB for inference \cite{chochlov2022using}. The ADA model, used in this work is available as a cloud service only \cite{openai}. Therefore the capability was added to SSCD to either use locally available models (CT5, CBF, GCB) for inference or to employ a cloud based OpenAI service (ADA) (see Figure~\ref{fig:sscd}).

SSCD was used in the following manner to evaluate individual-ANN inference and ensemble inference: given a combination of ANNs and source code, SSCD would execute clone detection with each model selected and capture the results, to assess individual ANN performance. Then, the clone candidates from each individual ANN-run would be merged to assess the combinations. For example, given a combination of CT5 and CBF (CT5\_CBF), SSCD would first execute with the CT5 model and then with the CBF model. For the ensemble, the resulting sets of clone candidates from each execution are merged into a final set of clone candidates, removing duplicates (resembling mathematical union operation). We hypothesize that this merging should drive higher recall, in that it aggregates the true-positives from both sets. But also important here is the merging effect on precision: intuitively the false positives in both sets, aggregated together, should negatively impact on precision and the extent of that impact is also of interest in a real-world context. Specifically, false positives increase the effort for developers who typically have to scan all the results of clone-detection techniques, for no clone-detection gain.

\section{Experimental Design}
\label{sec:methodology}
The objective of this experiment is to assess the effect of using different individual state-of-the-art ANNs, individually and as components of ensemble inference, towards improving the recall of clone detection over large-scale software systems, while preserving sufficient precision. This leads us to the following research question (RQs): 

\begin{enumerate}
    \item \textbf{How do differing ANNs affect the recall and overall efficacy of a large-scale, AI-based, clone-detection approach like SSCD?}
    \item \textbf{How does ensemble inference, based on combining these differing ANNs,  affect the recall and overall efficacy of a large-scale, AI-based, clone detection approach like SSCD?}
\end{enumerate}


Our industrial partners provided the C and C++ datasets, which we have made available online \cite{linkDataset} as resources for other researchers. The existing options for C/C++ benchmark datasets are already quite limited and those that do exist were deemed quite unreflective of code clones in realistic codebases by experts at our industrial partner company. Hence this new dataset was manually created by the company to provide realistic examples that spanned all four clone types.


From these, only clone pairs at function-level granularity are used, resulting in 70 such clone pairs for THE C benchmark and 83 clone pairs for C++ benchmark.

Standard metrics of recall, precision, and F-score \cite{chochlov2022using, Rattan2013} are used to assess the effectiveness of the individual ANNs and the subsequent ensemble inferencing, towards clone detection.

To answer the RQs, we assess how using ANN models (listed in Section~\ref{sec:background}), used in isolation and in combination compare to each other, focusing primarily on recall, but also considering precision and overall performance (the F-score). For example, given a combination CT5\_CBF, we can assess how its recall compares to CT5 and CBF when the latter two are used standalone, but also how it compares to ADA and GCB. We can also compare it to other ensembles to identify best-practice for developers. The exact steps to conduct this experiment are as follows:



\begin{enumerate}
    \item First, we run SSCD with both C/C++ datasets using the following parameters: minimum LOC = 0, search = naïve (faiss), code length = 128, no preprocessing (comments etc. retained), (similarity) threshold = 0, and topN = 10 (for parameters and their explanation please see below). In our previous work we found that the code length of 128 allows for effective and efficient neural network inference \cite{chochlov2022using}. The only variable parameter here is the choice of ANN used to compute the embedding for each code segment. We use four ANN models:\{ADA, CT5, CBF, GCB\} as described in Section~\ref{sec:background}. Therefore, 4 executions of SSCD are conducted: one for each individual model.
    \item In the second step, we obtain the best performing configuration, with regards to the F-score (given our requirements that improved recall should also have adequate precision). For this, we look at all possible configurations of similarity threshold and topN with given increase-steps, relying on the outputs obtained in the prior 4 executions. The increase-step for the similarity threshold is 0.01 and goes from 0 to 1 inclusively. The increase-step for topN is 1 and goes from 1 to 10 inclusively. Therefore, 100/10 threshold/topN combinations are inspected for each execution (obtained in Step 1). For example, for the ADA model, for the C++ benchmark, the best-performing configuration achieves 96.34\% F-score: this happens when the threshold is set to 0.91 and topN is set to 1. This is different for other models.
    \item Equipped with the reusults from these best-performing configurations, we combine them in all possible ways, resulting in 11 distinct combinations: \{ADA\_CT5, ADA\_CBF, ADA\_GCB, CT5\_CBF, CT5\_GCB, CBF\_GCB, ADA\_CT5\_CBF, ADA\_CT5\_GCB, ADA\_CBF\_GCB, CT5\_CBF\_GCB, ADA\_CT5\_CBF\_GCB\}.
    \item In the final step, the recall and F-score of the individual ANNs and their combinations are compared to assess their relative efficacy and the effectiveness of the ensembles.
\end{enumerate}

The meaning of the parameters to SSCD are:
\begin{itemize}
\item \textit{Minimum LOC}: The threshold number of lines of code that a C/C++ function must contain for SSCD to be considered as a clone candidate.
\item \textit{Search}: The efficient nearest neighbour library that is used to find similar embeddings. In this work we use the Faiss library from Facebook.
\item \textit{Code length}: The number of tokens of input source code from a given code segment that is used as input to the neural network when creating the embedding. Any subsequent tokens in the code segment are ignored.
\item \textit{Pre-processing}: SSCD offers several pre-processing choices to create a canonical form of the source code for each code fragment, such as rewriting the code to eliminate unnecessary white space. In the current paper we use no pre-processing.
\item \textit{Similarity threshold}: The nearest neighbour algorithm computes the cosine similarity between embeddings to find code segments with similar embeddings. Pairs of similar embeddings are the initial set of candidates that SSCD identifies as possible clones. Setting a similarity threshold causes SSCD to eliminate all candidate pairs with a similarity less than the threshold. Note that different neural networks can create neighbours with very different similarity scores. Thus, in the second step above, we search for a suitable similarity threshold for each of the four ANNs.
\item \textit{topN}: The nearest neighbour algorithm is configured to find the $topN$ nearest neighbours for each embedding. Finding a larger number of nearest neighbours for each embedding increases the execution time. The nearest neighbours for some embeddings may be very close, and for others very distant. Thus, the similarity threshold is used to eliminate neighbours that may be nearest to some embedding, but nonetheless distant.
\end{itemize}

\section{Results and Discussion}
\label{sec:results}

\subsection{Results}
The results of this experiment are presented in Table~\ref{tbl:cpp_results} and in Table~\ref{tbl:c_results} for C++ and C benchmarks respectively. The upper parts of these tables show recall, precision, and F-score for individual models and the lower parts show these statistics for models' combinations. In Table~\ref{tbl:c_results}, only 4 combinations are used: ADA is excluded because it achieves 100\% F-score individually and thus cannot be improved: any combination would only decrease precision.

\begin{table}[htbp]
\caption{Ensemble Inference Results for C++ Benchmark}
\label{tbl:cpp_results}
\centering
\resizebox{\columnwidth}{!}{\begin{tabular}{lrrr}
\hline
\multicolumn{1}{c}{\begin{tabular}[c]{@{}c@{}}Model/combination\\ name\end{tabular}} & \multicolumn{1}{c}{Recall (\%)} & \multicolumn{1}{c}{Precision (\%)} & \multicolumn{1}{c}{F-score (\%)} \\ \hline
\multicolumn{4}{l}{Individual models}                                                                                                                                                          \\ \hline
ADA                                                                                  & 95.18                           & 97.53                              & 96.34                            \\
CT5                                                                                  & 90.36                           & 96.15                              & 93.17                            \\
CBF                                                                                  & 81.93                           & 90.67                              & 86.08                            \\
GCB                                                                                  & 84.34                           & 89.74                              & 86.96                            \\ \hline
\multicolumn{4}{l}{Combinations}                                                                                                                                                               \\ \hline
ADA\_CT5                                                                             & \textbf{98.80}                  & 95.35                              & 97.04                            \\
ADA\_CBF                                                                             & \textbf{97.59}                  & 91.01                              & 94.19                            \\
ADA\_GCB                                                                             & \textbf{98.80}                  & 90.11                              & 94.26                            \\
CT5\_CBF                                                                             & \textbf{92.77}                  & 89.53                              & 91.12                            \\
CT5\_GCB                                                                             & \textbf{91.57}                  & 88.37                              & 89.94                            \\
CBF\_GCB                                                                             & \textbf{87.95}                  & 85.88                              & 86.90                            \\
ADA\_CT5\_CBF                                                                        & \textbf{98.80}                  & 89.13                              & 93.72                            \\
ADA\_CT5\_GCB                                                                        & \textbf{98.80}                  & 88.17                              & 93.18                            \\
ADA\_CBF\_GCB                                                                        & \textbf{98.80}                  & 86.32                              & 92.14                            \\
CT5\_CBF\_GCB                                                                        & \textbf{93.98}                  & 84.78                              & 89.14                            \\
ADA\_CT5\_CBF\_GCB                                                                   & \textbf{98.80}                  & 84.54                              & 91.12                            \\ \hline
\end{tabular}
}
\end{table}

\begin{table}[htbp]
\caption{Ensemble Inference Results for C Benchmark}
\label{tbl:c_results}
\centering
\resizebox{\columnwidth}{!}{\begin{tabular}{lrrr}
\hline
\multicolumn{1}{c}{\begin{tabular}[c]{@{}c@{}}Model/combination\\ name\end{tabular}} & \multicolumn{1}{c}{Recall (\%)} & \multicolumn{1}{c}{Precision (\%)} & \multicolumn{1}{c}{F-score (\%)} \\ \hline
\multicolumn{4}{l}{Individual models}                                                                                                                                                          \\ \hline
ADA                                                                                  & 100                             & 100                                & 100                              \\
CT5                                                                                  & 90                              & 92.65                              & 91.31                            \\
CBF                                                                                  & 77.14                           & 94.74                              & 85.04                            \\
GCB                                                                                  & 84.29                           & 90.77                              & 87.41                            \\ \hline
\multicolumn{4}{l}{Combinations}                                                                                                                                                               \\ \hline
CT5\_CBF                                                                             & \textbf{91.43}                  & 88.89                              & 90.14                            \\
CT5\_GCB                                                                             & \textbf{91.43}                  & 85.33                              & 88.27                            \\
CBF\_GCB                                                                             & \textbf{87.14}                  & 87.14                              & 87.14                            \\
CT5\_CBF\_GCB                                                                        & \textbf{92.86}                  & 82.28                              & 87.25                            \\ \hline
\end{tabular}
}
\end{table}

Overall the results show that ADA is the clear winner in terms of individual efficacy, both in terms of recall and precision. But the recall achieved during clone detection improved in all cases (15/15) when ensemble inference is compared to its component elements. For example, the ADA\_CT5 combination used on the C++ benchmark achieves 98.8\% recall, while, when used individually, the constituent models achieve only 95.18\% (ADA) and 90.36\% (CT5). 

For the ADA\_CT5 combination, the precision drops by 2.18\% compared to the ADA model and 0.8\% compared to the CT5 model. The worst-case precision drop, from the best individual model (ADA, 97.53\%) to the four-model combination ADA\_CT5\_CBF\_GCB (84.54\%), is a decrease of about 13.33\%. When we exclude ADA, the highest precision drop is from the individual CT5 model (96.15\%) to the three-model combination CT5\_CBF\_GCB (84.78\%), which is a decrease of approximately 11.82\%.

It is also important to note, that this recall (98.8\%) cannot be achieved by either ADA or CT5 individually, while maintaining the same precision (95.35\%): ADA cannot achieve this recall individually with any threshold/topN parameters in the threshold range of \{0 .. 1\} and the topN range of \{1 .. 10\}. In contrast, CT5 can achieve this recall (98.8\%) individually (the recall was manually increased to this number), but its precision and F-score decrease very markedly to 16.3\% and 27.98\% respectively. 

Looking at the results on the C benchmark, the individual ADA model reports perfect results, achieving maximum recall and precision. No combinations of ADA were considered as they would have only decreased precision for no possible recall gain. The best alternative individual model was CT5 achieving 90\% recall and 92.65 precision, but again there was a wide span in performance across the individual techniques. 

In terms of ensembles CT5\_CBF achieved a relatively high recall (91.43\%) with precision of 88.89\%. While a higher recall was achieved by CT5\_CBF\_GCB, that did come at the expense of precision, which decreased to 82.28\%.

The answers to the RQs posed above then are as follows:

\begin{enumerate}
    \item Differing ANNs significantly impact the recall and overall efficacy of SSCD-based clone detection, with ADA outperforming the other individual ANNs trialled;
    \item Ensemble inference improves recall further over the component ANNs (as expected) and, in the case of the ADA-CT5 ensemble, improves overall efficacy, as measured by the F-score. For other ensembles, while recall persistently improves, overall efficacy often deteriorates due to increased imprecision.
\end{enumerate}

\subsection{Discussion}
In terms of the individual ANNs, ADA outperforms the others significantly, particularly with respect to recall and overall F-score. CT5 is quite close in terms of precision but for best clone-detection performance, ADA is the most suitable candidate. 

However, ADA is a GPT-based model that is not publicly available. To use ADA we had to send each of our code fragments to the OpenAI platform, and pay a small dollar amount for ADA to generate the embeddings for each code fragment. In many cases, proprietary source code is a valuable business asset and, for such organizations, sending their source code to an external provider is unlikely to be acceptable.

If that is the case, then the best individual option is CT5: recall drops by 5\% on the C++ benchmark (10\% on the C benchmark) but code confidentiality is preserved. Alternatively the CT5\_CBF\_GCB ensemble could also be considered: using that approach improves recall to nearly 94\% on the C++ benchmark (92.86\% on the C benchmark) whilst ensuring confidentiality. Lower precision means the need to look at more false candidates. However, even though precision for this ensemble comes in as second-lowest across all approaches, a precision rate of 82-84\% does imply that 16-17 out of every 20 candidates proposed by the approach would be true positives, so not too much developer effort would be wasted during confirmation. A final point that needs to be made about this alternative is that all three approaches can be run in parallel (hardware resources permitting) meaning that speed performance is limited only by the slowest performing individual technique.

In terms of best-efficacy overall, regardless of inference-location, the choice for C is the individual ADA approach, albeit based on this admittedly small dataset.  For C++ the choice is slightly less clear: ADA offers the best precision but recall and overall F-score is is achieved using the ADA\_CT5 ensemble. 


\section{Threats to Validity}
\label{sec:threats}

We acknowledge ANN selection and sampling bias as one of the core threats to the validity of this experiment. Although, we've tried to select transformer-based ANNs with distinctive characteristics, those achieving state-of-the-art results, and those produced by different research teams (OpenAI - ADA; Microsoft - CBF, GCB; CT5- Salesforce), it is still possible that there are other existing models of this architecture-family (transformers) that would achieve better results and affect the ensemble results of this experiment by providing more diverse individual characteristics.


In addition, the sample size of of the code benchmarks is fairly small: it was approximately 480 KLOC in total with 153 clone, and was exclusively composed of C/C++ code. This suggests that the initial findings presented here should be buttressed by larger-scale studies and studies of different languages. 

But good large-scale datasets are exceptionally difficult to locate due to the nature of clone detection\cite{krinke2022bigclonebench}. For example, while the code employed here was scanned for additional clones, it is unlikely that the 153 clones identified as our gold-standard reflect all the clones in the dataset. Consequently out results may reflect a slightly higher recall than is absolutely correct and a slightly lower precision. To mitigate against this latter concern we manually inspected all 'false positives'  generated by the ANNs and found that all were indeed correctly categorized.

\section{Conclusions and Future Work}
\label{sec:conclusions}
In this work we employ differing state-of-the-art ANNs and ensembles of those ANNs to assess their effectiveness towards improving the recall of clone detection, with a view to improving large-scale clone detection. The results (see Section~\ref{sec:results}) suggest that ADA is the best individual technique and that typically ensemble inference can be used to improve the recall: it improved over the individual components in 15/15 cases assessed, although combinations of ADA were not considered for the C dataset because it had already achieved 100\% recall and precision in isolation. In this context, it's worth mentioning that the C dataset is relatively small, with just 80,190 lines of code (LOC), in contrast to the significantly larger C++ dataset, which contains 424,626 LOC. Thus the C++ dataset might have allowed for a more realistic evaluation.

There are several practical implications here:
\begin{itemize}
    \item In situations where sending source code to the cloud is acceptable, ADA seems like a promising alternative;
    \item Where ADA is not acceptable, CT5 is the best individual candidate;
    \item Ensemble inference can improve the recall of clone detection, if needed, and seems to have a low-cost practical application barrier;
    \item Ensemble inference can be helpful particularly if privacy and/or cost considerations exist. For example, the CT5\_CBF\_GCB combination achieves nearly as good recall (93.98\%) as ADA (a proprietary, paid-for service) (95.18\%) on the C++ dataset (see Table~\ref{tbl:cpp_results}). On the C dataset there is a larger distance between the approaches but the free, non-proprietary ensemble still identifies 92.86\% of the clones with acceptable precision.
    \item Inference time was not specifically assessed here but it depends on the execution design. When executed in parallel, the inference time becomes the maximum time of any ANNs in the ensemble. When executed sequentially, the inference time is a sum of all inference times of all the ANNs in the ensemble. 
\end{itemize}

Future work might include characterization of other ANN models and inspecting the impact of their characteristics towards improving the recall of clone detection (whilst preserving precision). Another direction might use larger benchmarks, such as the BCB\cite{Svajlenko2017a}, while remaining cognisant of the limitations of these larger datasets \cite{krinke2022bigclonebench}.

\section*{Acknowledgment}
This work was supported, in part, by Science Foundation Ireland grant 16/RC/3918 and by Huawei Technologies Co., Ltd.

\bibliographystyle{IEEEtran}
\bibliography{refs}

\end{document}